\documentclass[%
 aip,
 amsmath,amssymb,
 reprint,%
]{revtex4-1}

\usepackage{graphicx}
\usepackage{dcolumn}
\usepackage{bm}

\usepackage[utf8]{inputenc}
\usepackage[T1]{fontenc}
\usepackage{mathptmx}
\usepackage{multirow}
\usepackage{soul}

\begin{document}

\preprint{AIP/123-QED}

\title{Characterisation of Cryogenic Material Properties of 3D-Printed Superconducting Niobium using a 3D Lumped Element Microwave Cavity}

\author{Ben T. McAllister}
\affiliation{ARC Centre of Excellence for Engineered Quantum Systems, University of Western Australia, 35 Stirling Highway, Crawley WA 6009, Australia}\author{Jeremy Bourhill}
\affiliation{IMT Atlantique, Technopole Brest-Iroise, CS 83818, 29238 Brest Cedex 3, France andLab-STICC (UMR 6285), CNRS, Technopole Brest-Iroise, CS 83818, 29238 Brest Cedex 3, France}
\author{Wing Him J. Ma}
\affiliation{ARC Centre of Excellence for Engineered Quantum Systems, University of Western Australia, 35 Stirling Highway, Crawley WA 6009, Australia}
\author{Tim Sercombe}
\affiliation{School of Engineering, University of Western Australia, 35 Stirling Hwy, Crawley 6009, Australia}
\author{Maxim Goryachev}
\affiliation{ARC Centre of Excellence for Engineered Quantum Systems, University of Western Australia, 35 Stirling Highway, Crawley WA 6009, Australia}
\author{Michael E. Tobar}
\email{michael.tobar@uwa.edu.au}
\affiliation{ARC Centre of Excellence for Engineered Quantum Systems, University of Western Australia, 35 Stirling Highway, Crawley WA 6009, Australia}

\date{\today}

\begin{abstract}
We present an experimental characterisation of the electrical properties of 3D-printed Niobium. The study was performed by inserting a 3D-printed Nb post inside an Aluminium cylindrical cavity, forming a 3D lumped element re-entrant microwave cavity resonator. The resonator was cooled to temperatures below the critical temperature of Niobium (9.25K) and then Aluminium (1.2K), while measuring the quality factors of the electromagnetic resonances. This was then compared with finite element analysis of the cavity and a measurement of the same cavity with an Aluminium post of similar dimensions and frequency, to extract the surface resistance of the Niobium post. The 3D-printed Niobium exhibited a transition to the superconducting state at a similar temperature to the regular Niobium, as well as a surface resistance of $3.1\times10^{-4}$ $\Omega$. This value was comparable to many samples of traditionally machined Niobium previously studied without specialised surface treatment. Furthermore, this study demonstrates a simple new method for characterizing the material properties of a relatively small and geometrically simple sample of superconductor, which could be easily applied to other materials, particularly 3D-printed materials. Further research and development in additive manufacturing may see the application of 3D-printed Niobium in not only superconducting cavity designs, but in the innovative technology of the future. 
\end{abstract}

\maketitle

Devices made of superconducting materials are utilized in many areas of physics and engineering. For example, superconductors are applied in microwave oscillators, particle accelerators, gravitational wave detection, quantum technologies, and various tests of fundamental physics~\cite{Turneaure1968,Grimm05,tw,Lipa,nagel15,HSP2013a,HSP2013,HSP2011,jjar2010,Goryachev2019,Martinello2018,Cat19,3DLCAxions}, owing to their low-loss, zero DC electrical resistance, high resonant quality factors, and other properties. Embedding of superconducting qubits in superconducting resonant cavities has been the foundation for circuit quantum electrodynamics experiments and has markedly improved the coherence times of these qubits \cite{PaikPRL,reshitnyk2016}.

For use in these fields, the quality of production of these devices is set to a high standard. Niobium (Nb) has gained interest as a material of choice for superconducting devices due to its unique properties, such as the highest superconducting transition temperature amongst all pure metals (9.25 K \cite{Wong2011}). However, Nb especially is difficult to machine, and consequently the manufacturing of superconducting cavities and other devices, especially those with complex geometries, is expensive and time consuming \cite{Creedon2016}.

The growing field of additive manufacturing, also known as 3D-printing, has meant that the manufacturing of certain parts can be fast and inexpensive, and complex designs can be realised \cite{Gumbleton2019,Pellaton2018}. Also, waste material and labour are reduced when compared with traditional manufacturing. Additive manufacturing has in recent years been used to produce microwave resonator cavities for use in atomic clocks \cite{Affolderbach2018,Moreno2018} and ferromagnetic resonance spectroscopy \cite{castel,bourhill2019}. The high-precision and versatile nature of 3D-printing has been used to produce superconductors with complex geometries that are unattainable by conventional manufacturing techniques \cite{Holland2017,Wei2017,Creedon2016}. Clearly, 3D-printing has the potential to lead the manufacturing of superconducting cavities and other superconducting devices. This is of particular appeal for applications with Nb which, as stated, is difficult to machine by traditional methods. However, it is not generally known which properties of superconducting materials like Nb are retained when they are 3D printed. Nor if these materials will retain their superconductivity at low temperatures at all.

Consequently, this work aims to assess the feasibility of application of 3D-printed Nb to superconducting technology, by investigating the electrical and superconducting properties of a sample produced via Selective Laser Melting of Nb powder, by measuring the properties of electromagnetic modes in a resonant cavity containing the sample. This technique provides a powerful tool for determining material properties, and similar techniques (typically relying on the production of entire resonant cavities, or surfaces within those cavities) have been applied to a range of materials. A 3D-printed Nb rod was produced, and inserted into a microwave resonant cavity, forming a re-entrant cavity. This cavity was cooled to cryogenic temperatures, and the quality factors of the resonant modes were measured.  By comparing the measured values with those predicted from finite element modelling (FEM) of the system, we are able to extract material properties such as surface resistance.

This method of material characterization, using a physically small post constructed of the material under test, is a simple, inexpensive, and efficient means of characterizing material properties, and could be readily applied to a range of other materials, both 3D-printed and otherwise. Conveniently, this method requires only a relatively small and simple structure to be produced, as opposed to constructing an entire resonant cavity or some other more complex geometry.

\begin{figure}[h]
\includegraphics[width=0.4\textwidth]{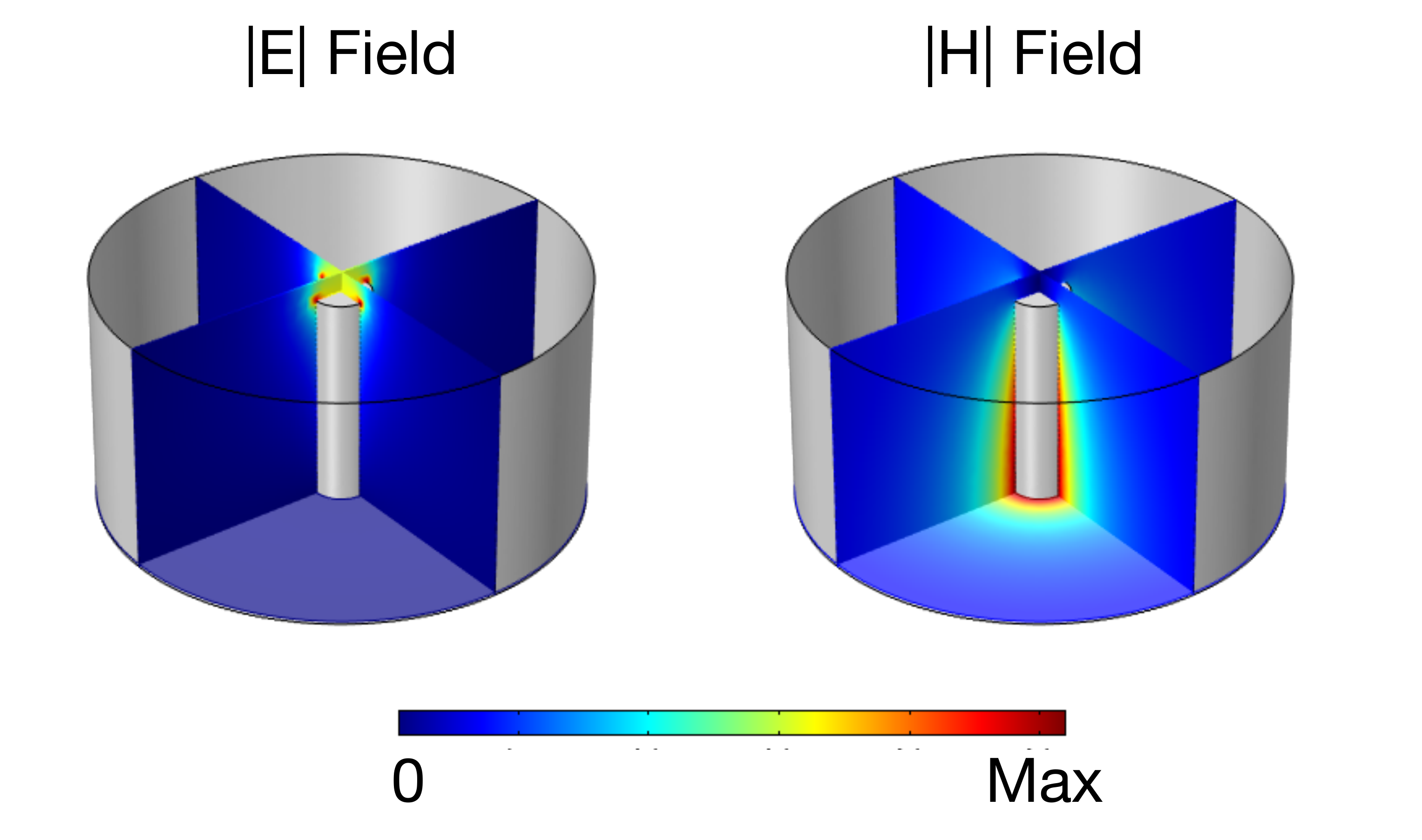}
\caption{Electric (left) and magnetic (right) field distributions in a re-entrant cavity. The cavity diameter was 16 mm, its height 8 mm, the diameter of the post is 2 mm and the gap spacing between the top of the post and the cavity roof is approximately 0.7 mm.}
\label{fig:cavfield}
\end{figure}

A re-entrant cavity is a hollow, conducting resonant cavity containing a conducting rod\cite{le2013rigorous,goryachev2015creating,Carvalho14,Barroso:2004aa}. A distinct feature of a re-entrant cavity is the existence of a gap between the top of the rod and the lid of the cavity. The cavity supports a series of electromagnetic resonant modes\cite{McAllisterPost,McAllisterP2}. The size of the gap between the post and the lid determines the capacitance of the effective lumped LC resonator, and hence strongly influences the resonant frequency of the cavity. For a given resonant re-entrant mode in the cavity, majority of the electrical energy is focused into this gap axially, whilst majority of the magnetic energy is concentrated radially about the post, in an azimuthal direction. Consequently, a re-entrant cavity can be viewed as an inductive post shorted with a lumped capacitor. An example of the field distributions in a re-entrant cavity is shown in figure~\ref{fig:cavfield}.

The intrinsic quality factor of a resonator, $Q_0$, describes the losses of the resonator. For an enclosed resonator, these losses are resistive, due to currents induced in the conducing walls. In the case of the re-entrant cavity, the total losses from the walls and post can be decomposed into,
\begin{equation}
    \frac{1}{Q_0}=\frac{1}{Q_{walls}}+\frac{1}{Q_{post}}.
    \label{eq:QLQI}
\end{equation} 
Here $Q_{walls}$ is inversely proportional to the losses in the walls of the cavity, and $Q_{post}$ is inversely proportional to the losses in the post.

There are other sources of loss, such as loading of the cavity by field-sensing probes $Q_{probes}$, and external losses $Q_{ext}$ due to, for example, small holes or gaps in the conducting walls of the resonator, and so any direct measurements of the quality factor of the resonator will always yield the loaded quality factor, $Q_L$, a parameter that incorporates all losses of the resonator. We can represent this as
\begin{equation}
    \frac{1}{Q_L}=\frac{1}{Q_{walls}}+\frac{1}{Q_{post}}+\frac{1}{Q_{probes}}+\frac{1}{Q_{ext}}.
    \label{eq:QLQI2}
\end{equation} 
In this study, we will use this relationship to determine the losses and material properties of the 3D-printed post.

The geometry factor of a resonance, $G$, denotes the ratio of the magnetic field in the volume of the cavity to the magnetic field on the surface. The geometry factor is defined as,
\begin{equation}
   G =\frac{\omega\mu_0 \iiint_V \mu_0|\vec{H}|^2 \,dV}{\iint_s|\vec{H_\tau}|^2\,dS},
   \label{eq:integral}
\end{equation}
where $\omega$ is the angular frequency of the mode, $\vec{H}$ is the magnetic field component of the mode, $\vec{H_\tau}$ the tangential magnetic field, and $\mu_0$ is the vacuum permeability. This parameter provides an avenue to examine the losses of a given mode, independent of the material. As stated, the primary loss mechanism in the conducting walls of a closed cavity is resistive losses due to currents induced by the oscillating magnetic field of the resonant mode. Consequently, the geometry factor is proportional to the intrinsic quality factor, and is related to $Q_0$ and the surface resistance of the material by

\begin{equation}
    Q_0=\frac{G}{R_s}.
    \label{eq:geometryfactor}
\end{equation}
Here $R_s$ is the surface resistance of the relevant material. Importantly, relationship~\eqref{eq:geometryfactor} holds assuming that the surfaces in the cavity all have the same surface resistance, but it can easily be extended such that each term contributing to $Q_0$, (e.g. $Q_{post}$, $Q_{walls}$ in this case) has its own relationship between $Q$, $G$, and $R_s$ as per~\eqref{eq:geometryfactor}. For example
\begin{equation}
    Q_0=\left(\frac{R_{post}}{G_{post}}+\frac{R_{walls}}{G_{walls}}\right) ^{-1},
    \label{eq:geometryfactor2}
\end{equation}
where the inverse relationship owes to the fact that losses add in parallel. Here $R_{post}$ is the surface resistance of the post, $R_{walls}$ is the surface resistance of the walls, $G_{post}$ is the geometry factor of the post and $G_{walls}$ is the geometry factor of the wall. Furthermore, assuming that the losses from probes and external factors are small (i.e. $Q_{probes}$, $Q_{ext}\gg Q_{post}$, $Q_{walls}$), the measured (or loaded) quality factor becomes very close to the intrinsic quality factor such that

\begin{equation}
   Q_L\approx Q_0\approx(\frac{R_{post}}{G_{post}}+\frac{R_{walls}}{G_{walls}})^{-1}.
    \label{eq:rseq}
\end{equation}
To compute the separate geometry factors, one simply computes~\eqref{eq:integral}, but with the surface $S$ defined as the relevant surface, either the post or walls of the cavity.

Now we can see that by measuring the loaded quality factor of a resonator such as a re-entrant cavity under the appropriate conditions, and computing geometry factors with FEM of the resonances, we can extract material properties such as $R_{post}$ and $R_{wall}$ with enough measurements.
\begin{figure}[t!]
\includegraphics[width=0.35\textwidth]{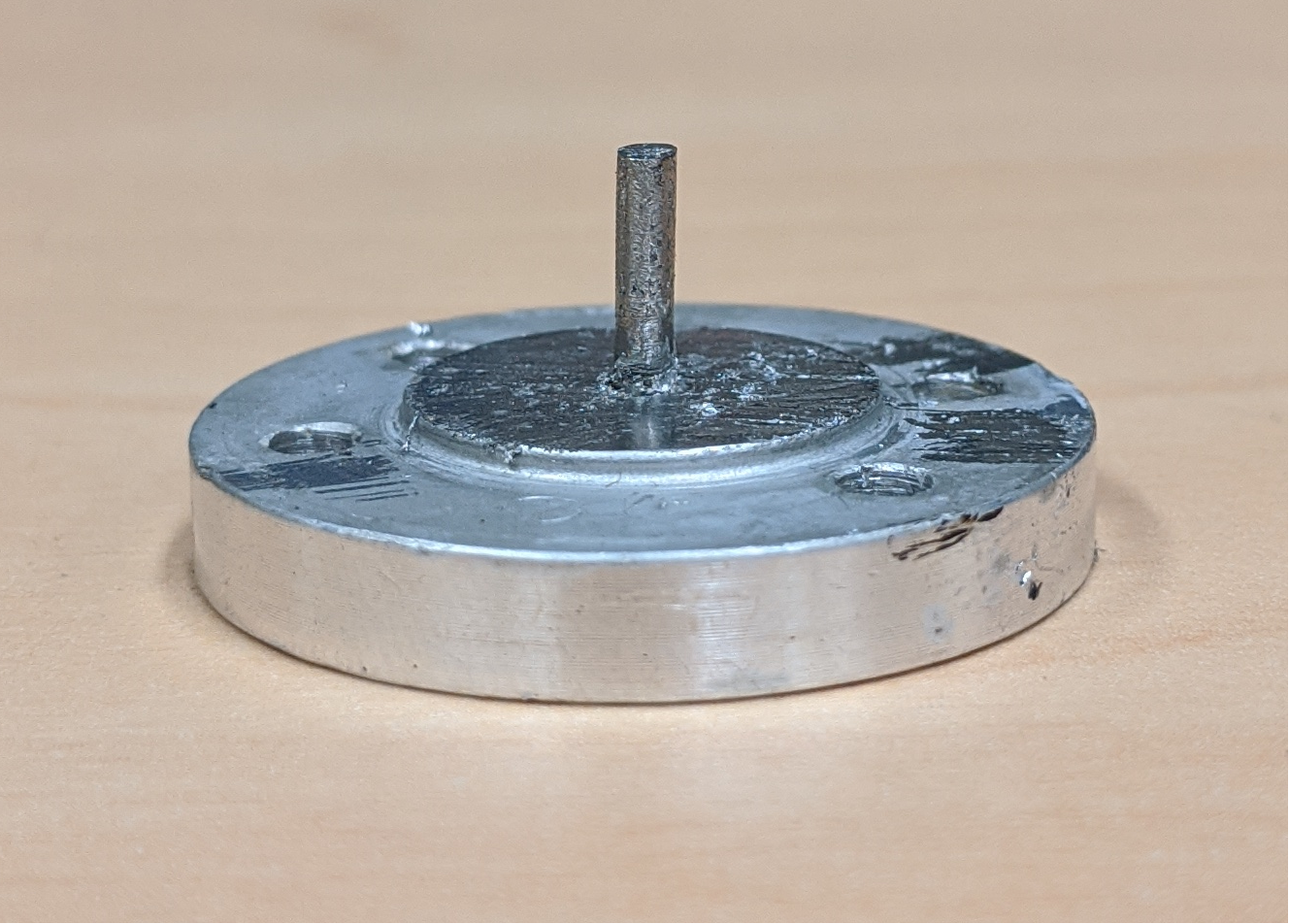}
\caption{Al cavity base with polished 3D-printed Nb post attached.}
\label{fig:post}
\end{figure}
The Nb post under test was produced using Selective Laser Melting (SLM), an additive manufacturing process, on a Realizer SLM100 using a laser power of 200 W, scan spacing of 100 $\mu$m, layer thickness of 50 $\mu$m and laser scan speed of 250 mm/s. The Nb (Nb) powder used to produce the post had a purity of 99.5 percent and had an average particle size of 34 $\mu$m. Before printing the chamber was purged with high purity Argon, and the Oxygen content was maintained at below 0.1 percent. Figure \ref{fig:post} shows two 3D-printed Nb posts viewed under a microscope, the polished post on the right was used for the experiment.

The post was soldered to the base of a cylindrical, pure Aluminium (Al) cavity to form the re-entrant cavity. The fundamental re-entrant mode was selected for study. FEM of the fields and cavity are presented in figure \ref{fig:cavfield}. The room temperature properties were measured with a Vector Network Analyser (VNA).

The probes were positioned such that their coupling parameters at room temperature were $< 10^{-4}$, guaranteeing that the losses due to probes would be small compared to material losses (ie $Q_{probes}\gg Q_{0}$), and~\eqref{eq:rseq} would hold.  The re-entrant cavity was installed inside a dilution refrigerator equipped with a 7 T superconducting magnet, which for this experiment served as a superconducting shield to isolate the experiment from ambient magnetic fields. The re-entrant cavity was connected again to a VNA, situated outside the refrigerator, with sufficient attenuation distributed along the input line to reduce the propagation of room temperature thermal noise into the experiment, and both cryogenic and room temperature amplifiers placed along the output line to ensure an appropriate signal to noise ratio for data anaylsis.

The re-entrant cavity was cooled and temperature stabilised at various cryogenic temperatures. At the base temperature, 20 mK, the frequency of the fundamental re-entrant mode was measured as 6.236 GHz. At each temperature, the power of the input signal was swept across a range of values and the transmitted power as a function of frequency was recorded.

In order to extract the material properties from~\eqref{eq:rseq}, more data was required. Given we have two unknown quantities in~\eqref{eq:rseq} for each measured $Q_L$ value at each temperature, ie $R_{post}$ (Nb) and $R_{walls}$ (Al), we require two sets of $Q_L$ measurements, which can then be turned into two simultaneous equations of the form~\eqref{eq:rseq} at each temperature.

In order to separate the Al losses from the Nb losses, and provide this second set of simultaneous equations, a new cavity base with a pure Al post with approximately the same height as the Nb post was manufactured. This base was attached to the cavity from the previous run. In this case, $R_{post}$ and $R_{walls}$ in~\eqref{eq:rseq} would be identical (both being the surface resistance of Al), allowing for their value to be acquired directly from the measured quality factor, $Q_L$, and geometry factors derived from FEM. This value for the surface resistance of Al at each temperature could then be substituted into the equations generated from the first run, with the Nb post, and used to solve for the Nb  surface resistance.

The same procedures of cooling the cavity in the dilution refrigerator, measuring transmitted power as a function of frequency and performing power dependence analysis were conducted on the second re-entrant cavity. The base temperature frequency was 6.351 GHz.

\begin{figure}[b!]
\includegraphics[width=0.43\textwidth]{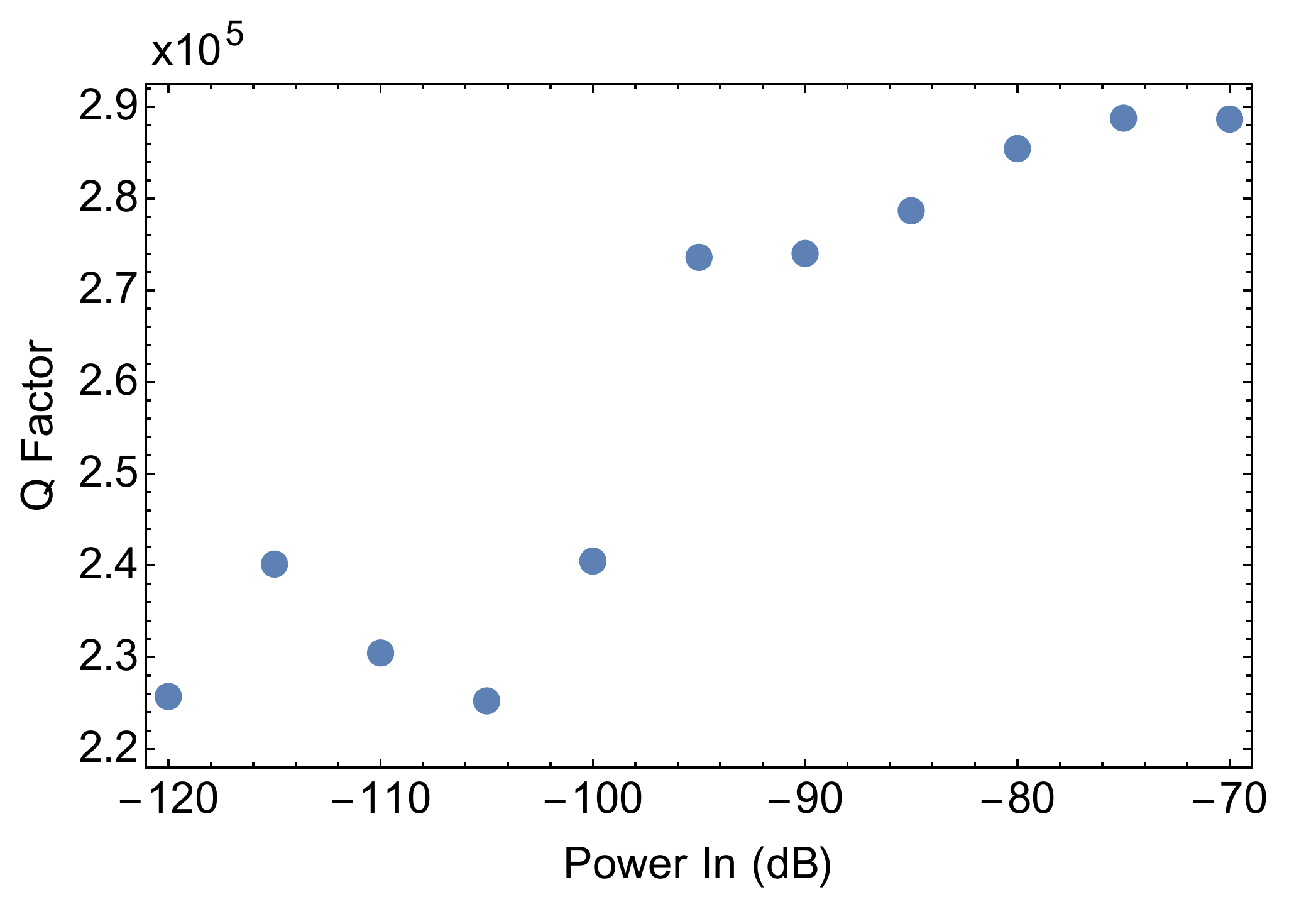}
\caption{Power dependence of quality factor at 20mK.}
\label{fig:powerdep}
\end{figure}

The data for both cavities at all temperatures and powers were fitted with the following equation to extract the quality factors: 
\begin{equation}
    P_{trans}=P_{inc}\frac{4\beta_1\beta_2}{(1+\beta_1+\beta_2)^2}\frac{1}{1+4Q^2_L(\frac{\omega-\omega_{res}}{\omega_{res}})^2}. 
    \label{powertransmittance}
\end{equation}

\begin{center}
\begin{table}[b]
\begin{tabular}{|r|c|c|c|c|c|c|c|}
\hline
\multicolumn{1}{|c|}{\multirow{2}{*}{Cavity}}& \multicolumn{1}{c|}{\multirow{1}{*}{Freq.}}& \multicolumn{1}{c|}{\multirow{1}{*}{$G_\text{post}$}} &\multicolumn{1}{c|}{ \multirow{1}{*}{$G_\text{wall}$}}& \multicolumn{4}{c|}{$Q_L$}\\ \cline{5-8}
& (GHz)&($\Omega$)&($\Omega$)&~300 K~& 20 K& 2 K& 0.02 K\\
\hline
Nb Post~&~6.236~&~123.13~&~157.32~&~396~&~1947~&~9292~&~288,730~\\
All Al~&~6.321~&~126.23~&~159.52~&~589~&~7000~&~7916~&~390,481~\\ \hline
\end{tabular}
\caption{Cavity parameters and $Q$ factors at various temperatures.}
\label{tab:1}
\end{table}
\end{center}

In equation \ref{powertransmittance}, $P_{inc}$ is the incident power, $P_{trans}$ is the transmitted power, $\omega_{res}$ is the resonant frequency of the mode and $\beta_1$ and $\beta_2$ are the coupling parameters of the input and output probes. The measured quality factors exhibited no obvious dependence on input power, other than at the base temperature of 20 mK. This is shown in figure \ref{fig:powerdep} for the Nb cavity, but a similar dependence was observed in the Al cavity. The degradation in the quality factor at low input power observed is due to the presence of a two level systems associated with a spurious loss mechanism. At low input power to the cavity, this mechanism absorbs energy and causes loss. At high input powers, the two level system is saturated and is no longer a source of spurious loss. At all temperatures in both cavities, the optimum measured quality factor was taken as the quality factor for analysis.

\begin{figure}[t!]
\includegraphics[width=0.5\textwidth]{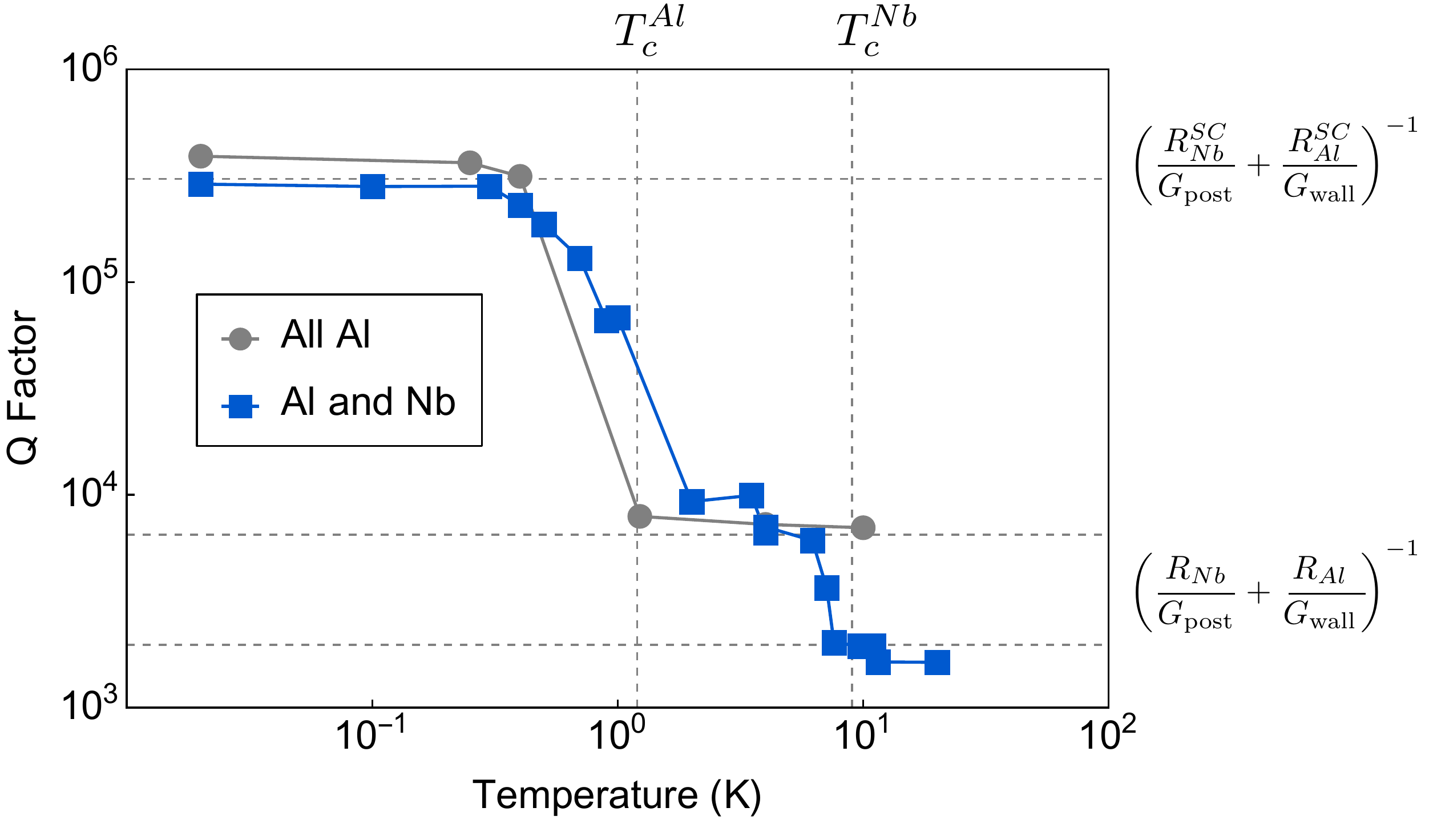}
\caption{Graph of quality factor versus temperature for both re-entrant cavities. The expressions for expected values of $Q_L$ on the right hand side are given as an example, to show how the simultaneous equations are constructed. These expressions relate to the Nb post cavity.}
\label{fig:NbQvsT}
\end{figure}

The extraction of quality factors at various temperatures for the two re-entrant cavities produced data sets of quality factor versus temperature, as shown in figure \ref{fig:NbQvsT}. The quality factor versus temperature for the re-entrant cavity with the Nb post displays two distinct ``jumps" in the quality factor. This is expected, owing to the fact that the cavity is composed of two distinct superconducting materials, with different transition temperatures. The jump at 7.6 K represents the Nb transition to the superconducting state. The divergence from the accepted value for this transition, $T_c^{Nb}=9.25$ K, can be attributed to fact that the temperature sensor in the dilution refrigerator is spatially separated from the Nb post, resulting in the temperature of Nb lagging that of the sensor during the cooldown procedure, only coming to thermal equilibrium after a stable temperature is reached and sustained. The jump at approximately 1 K in both figures matches closely with the well-known superconducting transition temperature of Al, $T_c^{Al}=1.2$ K.

To compute the geometry factors required for the extraction of $R_s$ values from these simultaneous equations, a re-entrant cavity model was constructed in COMSOL Multiphysics. A parametric sweep of the post height allowed the geometry factor of the walls and post to be determined for the two re-entrant cavities, which are at slightly different frequencies. The geometry factors, frequencies, and measure $Q_L$ values for both cavities are show in table 1.

Consequently,  the surface resistance of the 3D-printed Nb post, $R_{Nb}$ and the surface resistance of the pure Al, $R_{Al}$ could be determined at specific temperatures. With $G_\text{wall}$ and $G_\text{post}$ being calculated from FEM, $R_{Al}$ can be calculated from~\eqref{eq:rseq}, using the measured $Q_L$ values for the Al post cavity, given $R_\text{post}=R_\text{walls}$ in this case. Now, considering the Nb post cavity, the values of $R_{Al}$ at a given temperature can be substituted for $R_\text{walls}$ in (\ref{eq:rseq}). Using the measured $Q_L(T)$ values for the Nb post cavity (shown in figure \ref{fig:NbQvsT}), $R_\text{post}=R_{Nb}$ can be solved.

The resulting calculated resistance values for Al and Nb give $R_{Nb}^{SC}|_{20 mK}= 3.14\times10^{-4}~\Omega$, $R_{Al}^{SC}|_{20 mK}= 1.79\times10^{-4}~\Omega$ at the base temperature of 20 mK when both materials are in the superconducting state, and $R_{Nb}|_{10 K}$ = 0.0385$~\Omega$ and $R_{Al}|_{10 K}$= 0.0101$~\Omega$ in the normal conducting state, at 10 K. As expected, the surface resistances of the materials when they are superconducting are orders of magnitude less than when they are in the normal conducting state.

COMSOL modelling of small, empty cylindrical cavities yields low order TE and TM modes with GHz frequencies, and geometry factors on the order of $\sim$100 to $\sim$1000. If such cavities were to be constructed with the 3D-printed Nb studied here they would be expected to present quality factors on the order of $3\times 10^5$ to $3\times 10^6$, calculated using equation \ref{eq:geometryfactor} and the derived $R_{Nb}^{SC}$ value. Previously, we have measured untreated Nb resonant cavities to have quality factors ranging from $\sim 10^4$ to $\sim 10^7$. As the quality factor is mainly limited by the surface resistance, it can be deduced that the surface resistance of 3D-printed Nb is comparable, or in some cases even better than that of regular, untreated Nb. 
The consistency of the measured resistance values with previously observed Nb also demonstrates the value of the characterisation technique presented here. With only a small, 3D-printed Nb sample, we are able to extract the material properties with relative ease.

We have demonstrated a simple technique for characterizing the surface losses of materials. The method involves inserting a small, cylindrical sample of the material under test into a larger cylindrical cavity to form a re-entrant cavity, and then cooling the composite cavity to cryogenic temperatures. Through a combination of temperature and power dependent resonant quality factor measurements, and FEM, the surface resistances of the various materials in the cavity can be extracted from the experimental data. This technique has been applied to the characterization of a 3D-printed Nb post, finding a surface resistance value of $3.14\times10^{-4}~\Omega$ at 20 mK, and a superconducting transition temperature of 7.6 K. The measured value of surface resistance is consistent with previous measurements of non-3D-printed Nb, demonstrating the power of the technique, which may be applied to other materials. 

This work demonstrates that Nb can be 3D-printed without the loss of its appealing superconducting properties. This is promising, as Nb is a material of choice for superconducting systems in a range of physics and engineering applications, but it is notoriously difficult to machine. These measurements deal with untreated Nb - surface resistance measurements conducted on treated Nb have shown values of approximately 1n$\Omega$ \cite{Martinello2018}. Such treatments could be applied to 3D-printed Nb in future for lower surface resistance. 3D-printing of superconductors, and specifically Nb, has immense potential for future research, and may produce new applications which will be of great benefit in physics and engineering.

\begin{acknowledgments}
This work was funded by Australian Research Council grant number CE170100009. The data that support the findings of this study are available from the corresponding author upon reasonable request.
\end{acknowledgments}


\begin{thebibliography}{31}%
\makeatletter
\providecommand \@ifxundefined [1]{%
 \@ifx{#1\undefined}
}%
\providecommand \@ifnum [1]{%
 \ifnum #1\expandafter \@firstoftwo
 \else \expandafter \@secondoftwo
 \fi
}%
\providecommand \@ifx [1]{%
 \ifx #1\expandafter \@firstoftwo
 \else \expandafter \@secondoftwo
 \fi
}%
\providecommand \natexlab [1]{#1}%
\providecommand \enquote  [1]{``#1''}%
\providecommand \bibnamefont  [1]{#1}%
\providecommand \bibfnamefont [1]{#1}%
\providecommand \citenamefont [1]{#1}%
\providecommand \href@noop [0]{\@secondoftwo}%
\providecommand \href [0]{\begingroup \@sanitize@url \@href}%
\providecommand \@href[1]{\@@startlink{#1}\@@href}%
\providecommand \@@href[1]{\endgroup#1\@@endlink}%
\providecommand \@sanitize@url [0]{\catcode `\\12\catcode `\$12\catcode
  `\&12\catcode `\#12\catcode `\^12\catcode `\_12\catcode `\%12\relax}%
\providecommand \@@startlink[1]{}%
\providecommand \@@endlink[0]{}%
\providecommand \url  [0]{\begingroup\@sanitize@url \@url }%
\providecommand \@url [1]{\endgroup\@href {#1}{\urlprefix }}%
\providecommand \urlprefix  [0]{URL }%
\providecommand \Eprint [0]{\href }%
\providecommand \doibase [0]{http://dx.doi.org/}%
\providecommand \selectlanguage [0]{\@gobble}%
\providecommand \bibinfo  [0]{\@secondoftwo}%
\providecommand \bibfield  [0]{\@secondoftwo}%
\providecommand \translation [1]{[#1]}%
\providecommand \BibitemOpen [0]{}%
\providecommand \bibitemStop [0]{}%
\providecommand \bibitemNoStop [0]{.\EOS\space}%
\providecommand \EOS [0]{\spacefactor3000\relax}%
\providecommand \BibitemShut  [1]{\csname bibitem#1\endcsname}%
\let\auto@bib@innerbib\@empty
\bibitem [{\citenamefont {Turneaure}\ and\ \citenamefont
  {Weissman}(1968)}]{Turneaure1968}%
  \BibitemOpen
  \bibfield  {author} {\bibinfo {author} {\bibfnamefont {J.~P.}\ \bibnamefont
  {Turneaure}}\ and\ \bibinfo {author} {\bibfnamefont {I.}~\bibnamefont
  {Weissman}},\ }\bibfield  {title} {\enquote {\bibinfo {title} {Microwave
  surface resistance of superconducting niobium},}\ }\href {\doibase
  10.1063/1.1656986} {\bibfield  {journal} {\bibinfo  {journal} {Journal of
  Applied Physics}\ }\textbf {\bibinfo {volume} {39}},\ \bibinfo {pages}
  {4417--4427} (\bibinfo {year} {1968})}\BibitemShut {NoStop}%
\bibitem [{\citenamefont {Grimm}\ \emph {et~al.}(2005)\citenamefont {Grimm},
  \citenamefont {Aizaz}, \citenamefont {Johnson}, \citenamefont {Hartung},
  \citenamefont {Marti}, \citenamefont {Meidlinger}, \citenamefont
  {Meidlinger}, \citenamefont {Popielarski},\ and\ \citenamefont
  {York}}]{Grimm05}%
  \BibitemOpen
  \bibfield  {author} {\bibinfo {author} {\bibfnamefont {T.~L.}\ \bibnamefont
  {Grimm}}, \bibinfo {author} {\bibfnamefont {A.}~\bibnamefont {Aizaz}},
  \bibinfo {author} {\bibfnamefont {M.}~\bibnamefont {Johnson}}, \bibinfo
  {author} {\bibfnamefont {W.}~\bibnamefont {Hartung}}, \bibinfo {author}
  {\bibfnamefont {F.}~\bibnamefont {Marti}}, \bibinfo {author} {\bibfnamefont
  {D.}~\bibnamefont {Meidlinger}}, \bibinfo {author} {\bibfnamefont
  {M.}~\bibnamefont {Meidlinger}}, \bibinfo {author} {\bibfnamefont
  {J.}~\bibnamefont {Popielarski}}, \ and\ \bibinfo {author} {\bibfnamefont
  {R.~C.}\ \bibnamefont {York}},\ }\bibfield  {title} {\enquote {\bibinfo
  {title} {New directions in superconducting radio frequency cavities for
  accelerators},}\ }\href@noop {} {\bibfield  {journal} {\bibinfo  {journal}
  {IEEE Transactions on Applied Superconductivity}\ }\textbf {\bibinfo {volume}
  {15}},\ \bibinfo {pages} {2393--2396} (\bibinfo {year} {2005})}\BibitemShut
  {NoStop}%
\bibitem [{\citenamefont {Turneaure}\ \emph {et~al.}(1983)\citenamefont
  {Turneaure}, \citenamefont {Will}, \citenamefont {Farrell}, \citenamefont
  {Mattison},\ and\ \citenamefont {Vessot}}]{tw}%
  \BibitemOpen
  \bibfield  {author} {\bibinfo {author} {\bibfnamefont {J.~P.}\ \bibnamefont
  {Turneaure}}, \bibinfo {author} {\bibfnamefont {C.~M.}\ \bibnamefont {Will}},
  \bibinfo {author} {\bibfnamefont {B.~F.}\ \bibnamefont {Farrell}}, \bibinfo
  {author} {\bibfnamefont {E.~M.}\ \bibnamefont {Mattison}}, \ and\ \bibinfo
  {author} {\bibfnamefont {R.~F.~C.}\ \bibnamefont {Vessot}},\ }\bibfield
  {title} {\enquote {\bibinfo {title} {Test of the principle of equivalence by
  a null gravitational red-shift experiment},}\ }\href@noop {} {\bibfield
  {journal} {\bibinfo  {journal} {Phys. Rev. D}\ }\textbf {\bibinfo {volume}
  {27}},\ \bibinfo {pages} {1705--1714} (\bibinfo {year} {1983})}\BibitemShut
  {NoStop}%
\bibitem [{\citenamefont {Lipa}\ \emph {et~al.}(2003)\citenamefont {Lipa},
  \citenamefont {Nissen}, \citenamefont {Wang}, \citenamefont {Stricker},\ and\
  \citenamefont {Avaloff}}]{Lipa}%
  \BibitemOpen
  \bibfield  {author} {\bibinfo {author} {\bibfnamefont {J.~A.}\ \bibnamefont
  {Lipa}}, \bibinfo {author} {\bibfnamefont {J.~A.}\ \bibnamefont {Nissen}},
  \bibinfo {author} {\bibfnamefont {S.}~\bibnamefont {Wang}}, \bibinfo {author}
  {\bibfnamefont {D.~A.}\ \bibnamefont {Stricker}}, \ and\ \bibinfo {author}
  {\bibfnamefont {D.}~\bibnamefont {Avaloff}},\ }\bibfield  {title} {\enquote
  {\bibinfo {title} {New limit on signals of lorentz violation in
  electrodynamics},}\ }\href@noop {} {\bibfield  {journal} {\bibinfo  {journal}
  {Phys. Rev. Lett.}\ }\textbf {\bibinfo {volume} {90}},\ \bibinfo {pages}
  {060403} (\bibinfo {year} {2003})}\BibitemShut {NoStop}%
\bibitem [{\citenamefont {Nagel}\ \emph {et~al.}(2015)\citenamefont {Nagel},
  \citenamefont {Parker}, \citenamefont {Kovalchuk}, \citenamefont {Stanwix},
  \citenamefont {Hartnett}, \citenamefont {Ivanov}, \citenamefont {Peters},\
  and\ \citenamefont {Tobar}}]{nagel15}%
  \BibitemOpen
  \bibfield  {author} {\bibinfo {author} {\bibfnamefont {M.}~\bibnamefont
  {Nagel}}, \bibinfo {author} {\bibfnamefont {S.~R.}\ \bibnamefont {Parker}},
  \bibinfo {author} {\bibfnamefont {E.~V.}\ \bibnamefont {Kovalchuk}}, \bibinfo
  {author} {\bibfnamefont {P.~L.}\ \bibnamefont {Stanwix}}, \bibinfo {author}
  {\bibfnamefont {J.~G.}\ \bibnamefont {Hartnett}}, \bibinfo {author}
  {\bibfnamefont {E.~N.}\ \bibnamefont {Ivanov}}, \bibinfo {author}
  {\bibfnamefont {A.}~\bibnamefont {Peters}}, \ and\ \bibinfo {author}
  {\bibfnamefont {M.~E.}\ \bibnamefont {Tobar}},\ }\bibfield  {title} {\enquote
  {\bibinfo {title} {Direct terrestrial test of lorentz symmetry in
  electrodynamics to 10-18},}\ }\href@noop {} {\bibfield  {journal} {\bibinfo
  {journal} {Nat Commun}\ }\textbf {\bibinfo {volume} {6}},\ \bibinfo {pages}
  {10.1038/ncomms9174} (\bibinfo {year} {2015})}\BibitemShut {NoStop}%
\bibitem [{\citenamefont {Parker}\ \emph {et~al.}(2013)\citenamefont {Parker},
  \citenamefont {Hartnett}, \citenamefont {Povey},\ and\ \citenamefont
  {Tobar}}]{HSP2013a}%
  \BibitemOpen
  \bibfield  {author} {\bibinfo {author} {\bibfnamefont {S.~R.}\ \bibnamefont
  {Parker}}, \bibinfo {author} {\bibfnamefont {J.~G.}\ \bibnamefont
  {Hartnett}}, \bibinfo {author} {\bibfnamefont {R.~G.}\ \bibnamefont {Povey}},
  \ and\ \bibinfo {author} {\bibfnamefont {M.~E.}\ \bibnamefont {Tobar}},\
  }\bibfield  {title} {\enquote {\bibinfo {title} {Cryogenic resonant microwave
  cavity searches for hidden sector photons},}\ }\href@noop {} {\bibfield
  {journal} {\bibinfo  {journal} {Phys. Rev. D}\ }\textbf {\bibinfo {volume}
  {88}},\ \bibinfo {pages} {112004} (\bibinfo {year} {2013})}\BibitemShut
  {NoStop}%
\bibitem [{\citenamefont {Parker}, \citenamefont {Rybka},\ and\ \citenamefont
  {Tobar}(2013)}]{HSP2013}%
  \BibitemOpen
  \bibfield  {author} {\bibinfo {author} {\bibfnamefont {S.~R.}\ \bibnamefont
  {Parker}}, \bibinfo {author} {\bibfnamefont {G.}~\bibnamefont {Rybka}}, \
  and\ \bibinfo {author} {\bibfnamefont {M.~E.}\ \bibnamefont {Tobar}},\
  }\bibfield  {title} {\enquote {\bibinfo {title} {Hidden sector photon
  coupling of resonant cavities},}\ }\href@noop {} {\bibfield  {journal}
  {\bibinfo  {journal} {Phys. Rev. D}\ }\textbf {\bibinfo {volume} {87}},\
  \bibinfo {pages} {115008} (\bibinfo {year} {2013})}\BibitemShut {NoStop}%
\bibitem [{\citenamefont {Povey}, \citenamefont {Hartnett},\ and\ \citenamefont
  {Tobar}(2011)}]{HSP2011}%
  \BibitemOpen
  \bibfield  {author} {\bibinfo {author} {\bibfnamefont {R.~G.}\ \bibnamefont
  {Povey}}, \bibinfo {author} {\bibfnamefont {J.~G.}\ \bibnamefont {Hartnett}},
  \ and\ \bibinfo {author} {\bibfnamefont {M.~E.}\ \bibnamefont {Tobar}},\
  }\bibfield  {title} {\enquote {\bibinfo {title} {Microwave cavity hidden
  sector photon threshold crossing},}\ }\href@noop {} {\bibfield  {journal}
  {\bibinfo  {journal} {Phys. Rev. D}\ }\textbf {\bibinfo {volume} {84}},\
  \bibinfo {pages} {055023} (\bibinfo {year} {2011})}\BibitemShut {NoStop}%
\bibitem [{\citenamefont {Jaeckel}\ and\ \citenamefont
  {Ringwald}(2010)}]{jjar2010}%
  \BibitemOpen
  \bibfield  {author} {\bibinfo {author} {\bibfnamefont {J.}~\bibnamefont
  {Jaeckel}}\ and\ \bibinfo {author} {\bibfnamefont {A.}~\bibnamefont
  {Ringwald}},\ }\bibfield  {title} {\enquote {\bibinfo {title} {The low-energy
  frontier of particle physics},}\ }\href@noop {} {\bibfield  {journal}
  {\bibinfo  {journal} {Annual Review of Nuclear and Particle Science}\
  }\textbf {\bibinfo {volume} {60}},\ \bibinfo {pages} {405--437} (\bibinfo
  {year} {2010})}\BibitemShut {NoStop}%
\bibitem [{\citenamefont {Goryachev}, \citenamefont {McAllister},\ and\
  \citenamefont {Tobar}(2019)}]{Goryachev2019}%
  \BibitemOpen
  \bibfield  {author} {\bibinfo {author} {\bibfnamefont {M.}~\bibnamefont
  {Goryachev}}, \bibinfo {author} {\bibfnamefont {B.~T.}\ \bibnamefont
  {McAllister}}, \ and\ \bibinfo {author} {\bibfnamefont {M.~E.}\ \bibnamefont
  {Tobar}},\ }\bibfield  {title} {\enquote {\bibinfo {title} {Axion detection
  with precision frequency metrology},}\ }\href {\doibase
  10.1016/j.dark.2019.100345} {\bibfield  {journal} {\bibinfo  {journal}
  {Physics of the Dark Universe}\ }\textbf {\bibinfo {volume} {26}},\ \bibinfo
  {pages} {100345} (\bibinfo {year} {2019})}\BibitemShut {NoStop}%
\bibitem [{\citenamefont {Martinello}\ \emph {et~al.}(2018)\citenamefont
  {Martinello}, \citenamefont {Checchin}, \citenamefont {Romanenko},
  \citenamefont {Grassellino}, \citenamefont {Aderhold}, \citenamefont
  {Chandrasekeran}, \citenamefont {Melnychuk}, \citenamefont {Posen},\ and\
  \citenamefont {Sergatskov}}]{Martinello2018}%
  \BibitemOpen
  \bibfield  {author} {\bibinfo {author} {\bibfnamefont {M.}~\bibnamefont
  {Martinello}}, \bibinfo {author} {\bibfnamefont {M.}~\bibnamefont
  {Checchin}}, \bibinfo {author} {\bibfnamefont {A.}~\bibnamefont {Romanenko}},
  \bibinfo {author} {\bibfnamefont {A.}~\bibnamefont {Grassellino}}, \bibinfo
  {author} {\bibfnamefont {S.}~\bibnamefont {Aderhold}}, \bibinfo {author}
  {\bibfnamefont {S.~K.}\ \bibnamefont {Chandrasekeran}}, \bibinfo {author}
  {\bibfnamefont {O.}~\bibnamefont {Melnychuk}}, \bibinfo {author}
  {\bibfnamefont {S.}~\bibnamefont {Posen}}, \ and\ \bibinfo {author}
  {\bibfnamefont {D.~A.}\ \bibnamefont {Sergatskov}},\ }\bibfield  {title}
  {\enquote {\bibinfo {title} {Field-enhanced superconductivity in
  high-frequency niobium accelerating cavities},}\ }\href {\doibase
  10.1103/PhysRevLett.121.224801} {\bibfield  {journal} {\bibinfo  {journal}
  {Phys. Rev. Lett.}\ }\textbf {\bibinfo {volume} {121}},\ \bibinfo {pages}
  {224801} (\bibinfo {year} {2018})}\BibitemShut {NoStop}%
\bibitem [{\citenamefont {Thomson}\ \emph {et~al.}(2019)\citenamefont
  {Thomson}, \citenamefont {McAllister}, \citenamefont {Goryachev},
  \citenamefont {Ivanov},\ and\ \citenamefont {Tobar}}]{Cat19}%
  \BibitemOpen
  \bibfield  {author} {\bibinfo {author} {\bibfnamefont {C.~A.}\ \bibnamefont
  {Thomson}}, \bibinfo {author} {\bibfnamefont {B.~T.}\ \bibnamefont
  {McAllister}}, \bibinfo {author} {\bibfnamefont {M.}~\bibnamefont
  {Goryachev}}, \bibinfo {author} {\bibfnamefont {E.~N.}\ \bibnamefont
  {Ivanov}}, \ and\ \bibinfo {author} {\bibfnamefont {M.~E.}\ \bibnamefont
  {Tobar}},\ }\bibfield  {title} {\enquote {\bibinfo {title} {Results from
  upload-download: A phase-interferometric axion dark matter search},}\
  }\href@noop {} {\bibfield  {journal} {\bibinfo  {journal} {arXiv:1912.07751
  [hep-ex]}\ } (\bibinfo {year} {2019})}\BibitemShut {NoStop}%
\bibitem [{\citenamefont {McAllister}, \citenamefont {Parker},\ and\
  \citenamefont {Tobar}(2016)}]{3DLCAxions}%
  \BibitemOpen
  \bibfield  {author} {\bibinfo {author} {\bibfnamefont {B.~T.}\ \bibnamefont
  {McAllister}}, \bibinfo {author} {\bibfnamefont {S.~R.}\ \bibnamefont
  {Parker}}, \ and\ \bibinfo {author} {\bibfnamefont {M.~E.}\ \bibnamefont
  {Tobar}},\ }\bibfield  {title} {\enquote {\bibinfo {title} {{3D Lumped LC
  Resonators as Low Mass Axion Haloscopes}},}\ }\href {\doibase
  10.1103/PhysRevD.94.042001} {\bibfield  {journal} {\bibinfo  {journal} {Phys.
  Rev. D}\ }\textbf {\bibinfo {volume} {94}},\ \bibinfo {pages} {042001}
  (\bibinfo {year} {2016})},\ \Eprint {http://arxiv.org/abs/1605.05427}
  {arXiv:1605.05427 [physics.ins-det]} \BibitemShut {NoStop}%
\bibitem [{\citenamefont {Paik}\ \emph {et~al.}(2011)\citenamefont {Paik},
  \citenamefont {Schuster}, \citenamefont {Bishop}, \citenamefont {Kirchmair},
  \citenamefont {Catelani}, \citenamefont {Sears}, \citenamefont {Johnson},
  \citenamefont {Reagor}, \citenamefont {Frunzio}, \citenamefont {Glazman},
  \citenamefont {Girvin}, \citenamefont {Devoret},\ and\ \citenamefont
  {Schoelkopf}}]{PaikPRL}%
  \BibitemOpen
  \bibfield  {author} {\bibinfo {author} {\bibfnamefont {H.}~\bibnamefont
  {Paik}}, \bibinfo {author} {\bibfnamefont {D.~I.}\ \bibnamefont {Schuster}},
  \bibinfo {author} {\bibfnamefont {L.~S.}\ \bibnamefont {Bishop}}, \bibinfo
  {author} {\bibfnamefont {G.}~\bibnamefont {Kirchmair}}, \bibinfo {author}
  {\bibfnamefont {G.}~\bibnamefont {Catelani}}, \bibinfo {author}
  {\bibfnamefont {A.~P.}\ \bibnamefont {Sears}}, \bibinfo {author}
  {\bibfnamefont {B.~R.}\ \bibnamefont {Johnson}}, \bibinfo {author}
  {\bibfnamefont {M.~J.}\ \bibnamefont {Reagor}}, \bibinfo {author}
  {\bibfnamefont {L.}~\bibnamefont {Frunzio}}, \bibinfo {author} {\bibfnamefont
  {L.~I.}\ \bibnamefont {Glazman}}, \bibinfo {author} {\bibfnamefont {S.~M.}\
  \bibnamefont {Girvin}}, \bibinfo {author} {\bibfnamefont {M.~H.}\
  \bibnamefont {Devoret}}, \ and\ \bibinfo {author} {\bibfnamefont {R.~J.}\
  \bibnamefont {Schoelkopf}},\ }\bibfield  {title} {\enquote {\bibinfo {title}
  {Observation of high coherence in josephson junction qubits measured in a
  three-dimensional circuit qed architecture},}\ }\href@noop {} {\bibfield
  {journal} {\bibinfo  {journal} {Phys. Rev. Lett.}\ }\textbf {\bibinfo
  {volume} {107}},\ \bibinfo {pages} {240501} (\bibinfo {year}
  {2011})}\BibitemShut {NoStop}%
\bibitem [{\citenamefont {Reshitnyk}, \citenamefont {Jerger},\ and\
  \citenamefont {Fedorov}(2016)}]{reshitnyk2016}%
  \BibitemOpen
  \bibfield  {author} {\bibinfo {author} {\bibfnamefont {Y.}~\bibnamefont
  {Reshitnyk}}, \bibinfo {author} {\bibfnamefont {M.}~\bibnamefont {Jerger}}, \
  and\ \bibinfo {author} {\bibfnamefont {A.}~\bibnamefont {Fedorov}},\
  }\bibfield  {title} {\enquote {\bibinfo {title} {3d microwave cavity with
  magnetic flux control and enhanced quality factor},}\ }\href {\doibase
  10.1140/epjqt/s40507-016-0050-8} {\bibfield  {journal} {\bibinfo  {journal}
  {{EPJ} Quantum Technology}\ }\textbf {\bibinfo {volume} {3}} (\bibinfo {year}
  {2016}),\ 10.1140/epjqt/s40507-016-0050-8}\BibitemShut {NoStop}%
\bibitem [{\citenamefont {Wong}(2011)}]{Wong2011}%
  \BibitemOpen
  \bibfield  {author} {\bibinfo {author} {\bibfnamefont {T.~M.}\ \bibnamefont
  {Wong}},\ }\href@noop {} {\emph {\bibinfo {title} {Niobium: Properties,
  Production, and Applications}}}\ (\bibinfo  {publisher} {Nova Science
  Publishers},\ \bibinfo {year} {2011})\BibitemShut {NoStop}%
\bibitem [{\citenamefont {Creedon}\ \emph {et~al.}(2016)\citenamefont
  {Creedon}, \citenamefont {Goryachev}, \citenamefont {Kostylev}, \citenamefont
  {Sercombe},\ and\ \citenamefont {Tobar}}]{Creedon2016}%
  \BibitemOpen
  \bibfield  {author} {\bibinfo {author} {\bibfnamefont {D.~L.}\ \bibnamefont
  {Creedon}}, \bibinfo {author} {\bibfnamefont {M.}~\bibnamefont {Goryachev}},
  \bibinfo {author} {\bibfnamefont {N.}~\bibnamefont {Kostylev}}, \bibinfo
  {author} {\bibfnamefont {T.~B.}\ \bibnamefont {Sercombe}}, \ and\ \bibinfo
  {author} {\bibfnamefont {M.~E.}\ \bibnamefont {Tobar}},\ }\bibfield  {title}
  {\enquote {\bibinfo {title} {A 3d printed superconducting aluminium microwave
  cavity},}\ }\href@noop {} {\bibfield  {journal} {\bibinfo  {journal} {Applied
  Physics Letters}\ }\textbf {\bibinfo {volume} {109}},\ \bibinfo {pages}
  {032601} (\bibinfo {year} {2016})}\BibitemShut {NoStop}%
\bibitem [{\citenamefont {Gumbleton}\ \emph {et~al.}(2019)\citenamefont
  {Gumbleton}, \citenamefont {Cuenca}, \citenamefont {Klemencic}, \citenamefont
  {Jones},\ and\ \citenamefont {Porch}}]{Gumbleton2019}%
  \BibitemOpen
  \bibfield  {author} {\bibinfo {author} {\bibfnamefont {R.}~\bibnamefont
  {Gumbleton}}, \bibinfo {author} {\bibfnamefont {J.~A.}\ \bibnamefont
  {Cuenca}}, \bibinfo {author} {\bibfnamefont {G.~M.}\ \bibnamefont
  {Klemencic}}, \bibinfo {author} {\bibfnamefont {N.}~\bibnamefont {Jones}}, \
  and\ \bibinfo {author} {\bibfnamefont {A.}~\bibnamefont {Porch}},\ }\bibfield
   {title} {\enquote {\bibinfo {title} {Evaluating the coefficient of thermal
  expansion of additive manufactured alsi10mg using microwave techniques},}\
  }\href@noop {} {\bibfield  {journal} {\bibinfo  {journal} {Additive
  Manufacturing}\ }\textbf {\bibinfo {volume} {30}},\ \bibinfo {pages} {100841}
  (\bibinfo {year} {2019})}\BibitemShut {NoStop}%
\bibitem [{\citenamefont {Pellaton}\ \emph {et~al.}(2018)\citenamefont
  {Pellaton}, \citenamefont {Affolderbach}, \citenamefont {Skrivervik},
  \citenamefont {Ivanov}, \citenamefont {Debogovic}, \citenamefont {de~Rijk},\
  and\ \citenamefont {Mileti}}]{Pellaton2018}%
  \BibitemOpen
  \bibfield  {author} {\bibinfo {author} {\bibfnamefont {M.}~\bibnamefont
  {Pellaton}}, \bibinfo {author} {\bibfnamefont {C.}~\bibnamefont
  {Affolderbach}}, \bibinfo {author} {\bibfnamefont {A.}~\bibnamefont
  {Skrivervik}}, \bibinfo {author} {\bibfnamefont {A.}~\bibnamefont {Ivanov}},
  \bibinfo {author} {\bibfnamefont {T.}~\bibnamefont {Debogovic}}, \bibinfo
  {author} {\bibfnamefont {E.}~\bibnamefont {de~Rijk}}, \ and\ \bibinfo
  {author} {\bibfnamefont {G.}~\bibnamefont {Mileti}},\ }\bibfield  {title}
  {\enquote {\bibinfo {title} {3d printed microwave cavity for atomic clock
  applications: proof of concept},}\ }\href@noop {} {\bibfield  {journal}
  {\bibinfo  {journal} {Electronics Letters}\ }\textbf {\bibinfo {volume}
  {54}},\ \bibinfo {pages} {691--693} (\bibinfo {year} {2018})}\BibitemShut
  {NoStop}%
\bibitem [{\citenamefont {Affolderbach}\ \emph {et~al.}(2018)\citenamefont
  {Affolderbach}, \citenamefont {Moreno}, \citenamefont {Ivanov}, \citenamefont
  {Debogovic}, \citenamefont {Pellaton}, \citenamefont {Skrivervik},
  \citenamefont {De~Rijk},\ and\ \citenamefont {Mileti}}]{Affolderbach2018}%
  \BibitemOpen
  \bibfield  {author} {\bibinfo {author} {\bibfnamefont {C.}~\bibnamefont
  {Affolderbach}}, \bibinfo {author} {\bibfnamefont {W.}~\bibnamefont
  {Moreno}}, \bibinfo {author} {\bibfnamefont {A.}~\bibnamefont {Ivanov}},
  \bibinfo {author} {\bibfnamefont {T.}~\bibnamefont {Debogovic}}, \bibinfo
  {author} {\bibfnamefont {M.}~\bibnamefont {Pellaton}}, \bibinfo {author}
  {\bibfnamefont {A.}~\bibnamefont {Skrivervik}}, \bibinfo {author}
  {\bibfnamefont {E.}~\bibnamefont {De~Rijk}}, \ and\ \bibinfo {author}
  {\bibfnamefont {G.}~\bibnamefont {Mileti}},\ }\bibfield  {title} {\enquote
  {\bibinfo {title} {Study of additive manufactured microwave cavities for
  pulsed optically pumped atomic clock applications},}\ }\href@noop {}
  {\bibfield  {journal} {\bibinfo  {journal} {Applied Physics Letters}\
  }\textbf {\bibinfo {volume} {112}},\ \bibinfo {pages} {113502} (\bibinfo
  {year} {2018})}\BibitemShut {NoStop}%
\bibitem [{\citenamefont {Moreno}\ \emph {et~al.}(2018)\citenamefont {Moreno},
  \citenamefont {Affolderbach}, \citenamefont {Pellaton}, \citenamefont
  {Mileti}, \citenamefont {Ivanov}, \citenamefont {Skrivervik}, \citenamefont
  {Debogovic}, \citenamefont {Capdevila}, \citenamefont {Hoerni},\ and\
  \citenamefont {deRijk}}]{Moreno2018}%
  \BibitemOpen
  \bibfield  {author} {\bibinfo {author} {\bibfnamefont {W.}~\bibnamefont
  {Moreno}}, \bibinfo {author} {\bibfnamefont {C.}~\bibnamefont
  {Affolderbach}}, \bibinfo {author} {\bibfnamefont {M.}~\bibnamefont
  {Pellaton}}, \bibinfo {author} {\bibfnamefont {G.}~\bibnamefont {Mileti}},
  \bibinfo {author} {\bibfnamefont {A.}~\bibnamefont {Ivanov}}, \bibinfo
  {author} {\bibfnamefont {A.}~\bibnamefont {Skrivervik}}, \bibinfo {author}
  {\bibfnamefont {T.}~\bibnamefont {Debogovic}}, \bibinfo {author}
  {\bibfnamefont {S.}~\bibnamefont {Capdevila}}, \bibinfo {author}
  {\bibfnamefont {D.}~\bibnamefont {Hoerni}}, \ and\ \bibinfo {author}
  {\bibfnamefont {E.}~\bibnamefont {deRijk}},\ }\bibfield  {title} {\enquote
  {\bibinfo {title} {Ramsey-mode rb cell clock demonstration with a 3d-printed
  microwave cavity},}\ }in\ \href@noop {} {\emph {\bibinfo {booktitle} {2018
  European Frequency and Time Forum (EFTF)}}}\ (\bibinfo {organization}
  {IEEE},\ \bibinfo {year} {2018})\ pp.\ \bibinfo {pages} {75--79}\BibitemShut
  {NoStop}%
\bibitem [{\citenamefont {{Castel}}\ \emph {et~al.}(2019)\citenamefont
  {{Castel}}, \citenamefont {{Ammar}}, \citenamefont {{Manchec}}, \citenamefont
  {{Cochet}},\ and\ \citenamefont {{Youssef}}}]{castel}%
  \BibitemOpen
  \bibfield  {author} {\bibinfo {author} {\bibfnamefont {V.}~\bibnamefont
  {{Castel}}}, \bibinfo {author} {\bibfnamefont {S.~B.}\ \bibnamefont
  {{Ammar}}}, \bibinfo {author} {\bibfnamefont {A.}~\bibnamefont {{Manchec}}},
  \bibinfo {author} {\bibfnamefont {G.}~\bibnamefont {{Cochet}}}, \ and\
  \bibinfo {author} {\bibfnamefont {J.~B.}\ \bibnamefont {{Youssef}}},\
  }\bibfield  {title} {\enquote {\bibinfo {title} {Strong coupling of magnons
  to microwave photons in three-dimensional printed resonators},}\ }\href@noop
  {} {\bibfield  {journal} {\bibinfo  {journal} {IEEE Magnetics Letters}\
  }\textbf {\bibinfo {volume} {10}},\ \bibinfo {pages} {1--5} (\bibinfo {year}
  {2019})}\BibitemShut {NoStop}%
\bibitem [{\citenamefont {Bourhill}\ \emph {et~al.}(2019)\citenamefont
  {Bourhill}, \citenamefont {Castel}, \citenamefont {Manchec},\ and\
  \citenamefont {Cochet}}]{bourhill2019}%
  \BibitemOpen
  \bibfield  {author} {\bibinfo {author} {\bibfnamefont {J.}~\bibnamefont
  {Bourhill}}, \bibinfo {author} {\bibfnamefont {V.}~\bibnamefont {Castel}},
  \bibinfo {author} {\bibfnamefont {A.}~\bibnamefont {Manchec}}, \ and\
  \bibinfo {author} {\bibfnamefont {G.}~\bibnamefont {Cochet}},\ }\href@noop {}
  {\enquote {\bibinfo {title} {Spectroscopy of magnetic materials for universal
  characterisation of cavity-magnon polariton coupling strength},}\ } (\bibinfo
  {year} {2019}),\ \Eprint {http://arxiv.org/abs/1910.08333} {arXiv:1910.08333
  [physics.app-ph]} \BibitemShut {NoStop}%
\bibitem [{\citenamefont {Holland}\ \emph {et~al.}(2017)\citenamefont
  {Holland}, \citenamefont {Rosen}, \citenamefont {Materise}, \citenamefont
  {Woollett}, \citenamefont {Voisin}, \citenamefont {Wang}, \citenamefont
  {Torres}, \citenamefont {Mireles}, \citenamefont {Carosi},\ and\
  \citenamefont {DuBois}}]{Holland2017}%
  \BibitemOpen
  \bibfield  {author} {\bibinfo {author} {\bibfnamefont {E.~T.}\ \bibnamefont
  {Holland}}, \bibinfo {author} {\bibfnamefont {Y.~J.}\ \bibnamefont {Rosen}},
  \bibinfo {author} {\bibfnamefont {N.}~\bibnamefont {Materise}}, \bibinfo
  {author} {\bibfnamefont {N.}~\bibnamefont {Woollett}}, \bibinfo {author}
  {\bibfnamefont {T.}~\bibnamefont {Voisin}}, \bibinfo {author} {\bibfnamefont
  {Y.~M.}\ \bibnamefont {Wang}}, \bibinfo {author} {\bibfnamefont {S.~G.}\
  \bibnamefont {Torres}}, \bibinfo {author} {\bibfnamefont {J.}~\bibnamefont
  {Mireles}}, \bibinfo {author} {\bibfnamefont {G.}~\bibnamefont {Carosi}}, \
  and\ \bibinfo {author} {\bibfnamefont {J.~L.}\ \bibnamefont {DuBois}},\
  }\bibfield  {title} {\enquote {\bibinfo {title} {High-kinetic inductance
  additive manufactured superconducting microwave cavity},}\ }\href@noop {}
  {\bibfield  {journal} {\bibinfo  {journal} {Applied Physics Letters}\
  }\textbf {\bibinfo {volume} {111}},\ \bibinfo {pages} {202602} (\bibinfo
  {year} {2017})}\BibitemShut {NoStop}%
\bibitem [{\citenamefont {Wei}\ \emph {et~al.}(2017)\citenamefont {Wei},
  \citenamefont {Peng}, \citenamefont {Xie}, \citenamefont {Xue}, \citenamefont
  {Wang},\ and\ \citenamefont {Ding}}]{Wei2017}%
  \BibitemOpen
  \bibfield  {author} {\bibinfo {author} {\bibfnamefont {X.}~\bibnamefont
  {Wei}}, \bibinfo {author} {\bibfnamefont {E.}~\bibnamefont {Peng}}, \bibinfo
  {author} {\bibfnamefont {Y.}~\bibnamefont {Xie}}, \bibinfo {author}
  {\bibfnamefont {J.}~\bibnamefont {Xue}}, \bibinfo {author} {\bibfnamefont
  {J.}~\bibnamefont {Wang}}, \ and\ \bibinfo {author} {\bibfnamefont
  {J.}~\bibnamefont {Ding}},\ }\bibfield  {title} {\enquote {\bibinfo {title}
  {Extrusion printing of a designed three-dimensional yba2cu3o7- x
  superconductor with milled precursor powder},}\ }\href@noop {} {\bibfield
  {journal} {\bibinfo  {journal} {Journal of Materials Chemistry C}\ }\textbf
  {\bibinfo {volume} {5}},\ \bibinfo {pages} {3382--3389} (\bibinfo {year}
  {2017})}\BibitemShut {NoStop}%
\bibitem [{\citenamefont {Le~Floch}\ \emph {et~al.}(2013)\citenamefont
  {Le~Floch}, \citenamefont {Fan}, \citenamefont {Aubourg}, \citenamefont
  {Cros}, \citenamefont {Carvalho}, \citenamefont {Shan}, \citenamefont
  {Bourhill}, \citenamefont {Ivanov}, \citenamefont {Humbert}, \citenamefont
  {Madrangeas} \emph {et~al.}}]{le2013rigorous}%
  \BibitemOpen
  \bibfield  {author} {\bibinfo {author} {\bibfnamefont {J.-M.}\ \bibnamefont
  {Le~Floch}}, \bibinfo {author} {\bibfnamefont {Y.}~\bibnamefont {Fan}},
  \bibinfo {author} {\bibfnamefont {M.}~\bibnamefont {Aubourg}}, \bibinfo
  {author} {\bibfnamefont {D.}~\bibnamefont {Cros}}, \bibinfo {author}
  {\bibfnamefont {N.}~\bibnamefont {Carvalho}}, \bibinfo {author}
  {\bibfnamefont {Q.}~\bibnamefont {Shan}}, \bibinfo {author} {\bibfnamefont
  {J.}~\bibnamefont {Bourhill}}, \bibinfo {author} {\bibfnamefont
  {E.}~\bibnamefont {Ivanov}}, \bibinfo {author} {\bibfnamefont
  {G.}~\bibnamefont {Humbert}}, \bibinfo {author} {\bibfnamefont
  {V.}~\bibnamefont {Madrangeas}},  \emph {et~al.},\ }\bibfield  {title}
  {\enquote {\bibinfo {title} {Rigorous analysis of highly tunable cylindrical
  transverse magnetic mode re-entrant cavities},}\ }\href@noop {} {\bibfield
  {journal} {\bibinfo  {journal} {Review of Scientific Instruments}\ }\textbf
  {\bibinfo {volume} {84}},\ \bibinfo {pages} {125114} (\bibinfo {year}
  {2013})}\BibitemShut {NoStop}%
\bibitem [{\citenamefont {Goryachev}\ and\ \citenamefont
  {Tobar}(2015)}]{goryachev2015creating}%
  \BibitemOpen
  \bibfield  {author} {\bibinfo {author} {\bibfnamefont {M.}~\bibnamefont
  {Goryachev}}\ and\ \bibinfo {author} {\bibfnamefont {M.~E.}\ \bibnamefont
  {Tobar}},\ }\bibfield  {title} {\enquote {\bibinfo {title} {Creating tuneable
  microwave media from a two-dimensional lattice of re-entrant posts},}\
  }\href@noop {} {\bibfield  {journal} {\bibinfo  {journal} {Journal of Applied
  Physics}\ }\textbf {\bibinfo {volume} {118}},\ \bibinfo {pages} {204504}
  (\bibinfo {year} {2015})}\BibitemShut {NoStop}%
\bibitem [{\citenamefont {Carvalho}\ \emph {et~al.}(2014)\citenamefont
  {Carvalho}, \citenamefont {Fan}, \citenamefont {Le~Floch},\ and\
  \citenamefont {Tobar}}]{Carvalho14}%
  \BibitemOpen
  \bibfield  {author} {\bibinfo {author} {\bibfnamefont {N.~C.}\ \bibnamefont
  {Carvalho}}, \bibinfo {author} {\bibfnamefont {Y.}~\bibnamefont {Fan}},
  \bibinfo {author} {\bibfnamefont {J.-M.}\ \bibnamefont {Le~Floch}}, \ and\
  \bibinfo {author} {\bibfnamefont {M.~E.}\ \bibnamefont {Tobar}},\ }\bibfield
  {title} {\enquote {\bibinfo {title} {Piezoelectric voltage coupled reentrant
  cavity resonator},}\ }\href {\doibase 10.1063/1.4897482} {\bibfield
  {journal} {\bibinfo  {journal} {Review of Scientific Instruments}\ }\textbf
  {\bibinfo {volume} {85}},\ \bibinfo {pages} {104705} (\bibinfo {year}
  {2014})},\ \Eprint {http://arxiv.org/abs/https://doi.org/10.1063/1.4897482}
  {https://doi.org/10.1063/1.4897482} \BibitemShut {NoStop}%
\bibitem [{\citenamefont {Barroso}\ \emph {et~al.}(2004)\citenamefont
  {Barroso}, \citenamefont {Castro}, \citenamefont {Aguiar},\ and\
  \citenamefont {Carneiro}}]{Barroso:2004aa}%
  \BibitemOpen
  \bibfield  {author} {\bibinfo {author} {\bibfnamefont {J.~J.}\ \bibnamefont
  {Barroso}}, \bibinfo {author} {\bibfnamefont {P.~J.}\ \bibnamefont {Castro}},
  \bibinfo {author} {\bibfnamefont {O.~D.}\ \bibnamefont {Aguiar}}, \ and\
  \bibinfo {author} {\bibfnamefont {L.~A.}\ \bibnamefont {Carneiro}},\
  }\bibfield  {title} {\enquote {\bibinfo {title} {Reentrant cavities as
  electromechanical transducers},}\ }\bibfield  {booktitle} {\emph {\bibinfo
  {booktitle} {Review of Scientific Instruments}},\ }\href {\doibase
  10.1063/1.1688438} {\bibfield  {journal} {\bibinfo  {journal} {Review of
  Scientific Instruments}\ }\textbf {\bibinfo {volume} {75}},\ \bibinfo {pages}
  {1000--1005} (\bibinfo {year} {2004})}\BibitemShut {NoStop}%
\bibitem [{\citenamefont {McAllister}\ \emph {et~al.}(2017)\citenamefont
  {McAllister}, \citenamefont {Shen}, \citenamefont {Flower}, \citenamefont
  {Parker},\ and\ \citenamefont {Tobar}}]{McAllisterPost}%
  \BibitemOpen
  \bibfield  {author} {\bibinfo {author} {\bibfnamefont {B.~T.}\ \bibnamefont
  {McAllister}}, \bibinfo {author} {\bibfnamefont {Y.}~\bibnamefont {Shen}},
  \bibinfo {author} {\bibfnamefont {G.}~\bibnamefont {Flower}}, \bibinfo
  {author} {\bibfnamefont {S.~R.}\ \bibnamefont {Parker}}, \ and\ \bibinfo
  {author} {\bibfnamefont {M.~E.}\ \bibnamefont {Tobar}},\ }\bibfield  {title}
  {\enquote {\bibinfo {title} {Higher order reentrant post modes in cylindrical
  cavities},}\ }\href {\doibase 10.1063/1.4991751} {\bibfield  {journal}
  {\bibinfo  {journal} {Journal of Applied Physics}\ }\textbf {\bibinfo
  {volume} {122}},\ \bibinfo {pages} {144501} (\bibinfo {year} {2017})},\
  \Eprint {http://arxiv.org/abs/https://doi.org/10.1063/1.4991751}
  {https://doi.org/10.1063/1.4991751} \BibitemShut {NoStop}%
\bibitem [{\citenamefont {McAllister}\ and\ \citenamefont
  {Tobar}(2018)}]{McAllisterP2}%
  \BibitemOpen
  \bibfield  {author} {\bibinfo {author} {\bibfnamefont {B.~T.}\ \bibnamefont
  {McAllister}}\ and\ \bibinfo {author} {\bibfnamefont {M.~E.}\ \bibnamefont
  {Tobar}},\ }\bibfield  {title} {\enquote {\bibinfo {title} {Response to
  ``comment on `higher order reentrant post modes in cylindrical cavities'''
  [j. appl. phys. 123, 226101 (2018)]},}\ }\href {\doibase 10.1063/1.5024564}
  {\bibfield  {journal} {\bibinfo  {journal} {Journal of Applied Physics}\
  }\textbf {\bibinfo {volume} {123}},\ \bibinfo {pages} {226102} (\bibinfo
  {year} {2018})},\ \Eprint
  {http://arxiv.org/abs/https://doi.org/10.1063/1.5024564}
  {https://doi.org/10.1063/1.5024564} \BibitemShut {NoStop}%
\end{thebibliography}
\end{document}